\documentclass[epj]{svjour}

\usepackage{graphicx}
\usepackage{fancyhdr}
\graphicspath{{figs/}}
\def\makeheadbox{{
 }}
\usepackage{amsfonts}
\usepackage{amsmath}
\usepackage{graphicx} 
\usepackage{amssymb}

\def\beq{\begin{equation}}
\def\eeq{\end{equation}}
\def\nn{\nonumber}
\def\bea{\begin{eqnarray}}
\def\eea{\end{eqnarray}}
\def\ba{\begin{array}}                  
\def\ea{\end{array}}

\newcommand{\br}{{\rm BR}}
\newcommand{\amu}{a_\mu}
\newcommand{\amuexp}{a_\mu^{\rm exp}}
\newcommand{\amutheo}{a_\mu^{\rm theo}}

\def\mytitle{My title} 
\def\myauthors{My name}  
\def\mytype{My type of session}
\def\mysession{My session}

\def\mytitle{Finite Unified Theories} 
\def\myauthors{Myriam Mondrag\'on}    
\def\mytype{Contributed Talk}    
\def\mysession{Theoretical Models}

\begin{document}
\title{Finite Unified Theories confronted with low-energy phenomenology}
\author{S. Heinemeyer\inst{1}
 \and
 M. Mondrag\'on\inst{2}
\thanks{\emph{Email:}myriam@fisica.unam.mx}%
 \and
 G. Zoupanos\inst{3}
}                     
%
%
\institute{Instituto de Fisica de Cantabria (CSIC-UC), Santander,
  Spain
\and Instituto de F\'{\i}sica,
             Universidad Nacional Aut\'onoma de M\'exico,
              M\'exico
\and Physics Department, National Technical University,
Athens, Greece
}
%
\date{}
\abstract{ Finite Unified Theories (FUTs) are N=1 supersymmetric Grand
  Unified Theories that can be made all-loop finite.  The requirement
  of all-loop finiteness leads to a severe reduction of the free
  parameters of the theory and, in turn, to a large number of
  predictions.  Here SU(5) FUTs are investigated in the context of
  low-energy phenomenology observables.  We present a detailed
  scanning of these FUTs, including theoretical uncertainties at the
  unification scale and applying all phenomenological constraints.
  Taking into account the restrictions from the top and
  bottom quark masses, we can discriminate
  between different models. Including further low-energy constraints
  such as $B$~physics   observables, the bound on the lightest Higgs
  boson mass and the cold dark matter density, we
  determine the predictions of
  the allowed parameter space for the Higgs boson sector and the
  supersymmetric particle spectrum of the model.
} 
\maketitle
%

\section{Introduction}

Finite Unified Theories are $N=1$ supersymmetric Grand Unified
Theories (GUTs) which can be made finite  to all-loop orders,
including the soft supersymmetry breaking sector.  To
construct GUTs with reduced independent parameters
\cite{Kubo:1994bj,zoup-zim1} one has to search for renormalization
group invariant (RGI) relations holding below the Planck scale, which
in turn are preserved down to the GUT scale. Of particular interest is
the possibility to find RGI relations among couplings that guarantee
finitenes to all-orders in perturbation theory
\cite{zoup-lucchesi1,zoup-ermushev1}.  In order to achieve the latter
it is enough to study the uniqueness of the solutions to the one-loop
finiteness conditions \cite{zoup-lucchesi1,zoup-ermushev1}.  The
constructed {\it finite unified} $N=1$ supersymmetric SU(5) GUTs using
the above tools, predicted correctly from the dimensionless sector
(Gauge-Yukawa unification), among others, the top quark mass
\cite{zoup-finite1}.  The search for RGI relations and finiteness has
been extended to the soft supersymmetry breaking sector (SSB) of these
theories \cite{zoup-kmz2,zoup-jack2}, which involves parameters of
dimension one and two.  Eventually, the full theories can be made
all-loop finite and their predictive power is extended to the Higgs
sector and the s-spectrum. 

Finiteness can be understood by considering a chiral, anomaly free,
$N=1$ globally supersymmetric
gauge theory based on a group G with gauge coupling
constant $g$. The
superpotential of the theory is given by
\begin{equation}
 W= \frac{1}{2}\,m^{ij} \,\Phi_{i}\,\Phi_{j}+
\frac{1}{6}\,C^{ijk} \,\Phi_{i}\,\Phi_{j}\,\Phi_{k}~, 
\label{1}
\end{equation}
where $m^{ij}$ (the mass terms) and $C^{ijk}$ (the Yukawa couplings) are
gauge invariant tensors and 
the matter field $\Phi_{i}$ transforms
according to the irreducible representation  $R_{i}$
of the gauge group $G$. 
All the one-loop $\beta$-functions of the theory
vanish if  the $\beta$-function of the gauge coupling $\beta_g^{(1)}$, and
the anomalous dimensions of the Yukawa couplings $\gamma_i^{j(1)}$, vanish, i.e.
\begin{equation}
\sum _i \ell (R_i) = 3 C_2(G) \,,~
\frac{1}{2}C_{ipq} C^{jpq} = 2\delta _i^j g^2  C_2(R_i)\ ,
\label{2}
\end{equation}
where $\ell (R_i)$ is the Dynkin index of $R_i$, and $C_2(G)$ is the
quadratic Casimir invariant of the adjoint 
representation of $G$. 
A theorem \cite{zoup-lucchesi1} guarantees
the vanishing of the $\beta$-functions to all-orders in perturbation
theory.  This requires that, in addition to the
one-loop finiteness conditions (\ref{2}),  
the Yukawa couplings are reduced
in favour of the gauge coupling.
Alternatively, similar results can be
obtained  \cite{zoup-ermushev1,zoup-strassler} using an analysis of
the all-loop NSVZ gauge beta-function \cite{zoup-novikov1}. 

In the soft breaking sector, it was found that RGI SSB scalar masses
in Gauge-Yukawa unified models satisfy a universal sum rule at
one-loop \cite{zoup-kkk1}. This result was generalized to two-loops
for finite theories \cite{zoup-kkmz1}, and then to all-loops for
general Gauge-Yukawa and finite unified theories \cite{zoup-kkz}.
Then the following soft scalar-mass sum rule is found
\cite{zoup-kkmz1}
\begin{equation}
\frac{(~m_{i}^{2}+m_{j}^{2}+m_{k}^{2}~)}{M M^{\dag}} =
1+\frac{g^2}{16 \pi^2}\,\Delta^{(1)}
+O(g^4)~
\label{zoup-sumr}
\end{equation}
for i, j, k with $\rho^{ijk}_{(0)} \neq 0$, where $\Delta^{(1)}$ is
the two-loop correction
\begin{equation}
\Delta^{(1)} =  -2\sum_{l} [(m^{2}_{l}/M M^{\dag})-(1/3)]~ \ell (R_l),
\label{5}
\end{equation}
$\Delta^{(1)}$ vanishes for the
universal choice, i.e.\ when all the soft scalar masses are the same at
the unification point.

\section{ FINITE UNIFIED THEORIES}

The first one- and two-loop finite model was presented in
\cite{Jones:1984qd}. 
A predictive Gauge-Yukawa unified $SU(5)$ model which is finite to all
orders, in addition to the requirements mentioned already, should also
have the following properties:
\begin{enumerate}
\item 
One-loop anomalous dimensions are diagonal,
i.e.,  $\gamma_{i}^{(1)\,j} \propto \delta^{j}_{i} $.
\item Three fermion generations, in the irreducible representations
  $\overline{\bf 5}_{i},{\bf 10}_i~(i=1,2,3)$, which obviously should
  not couple to the adjoint ${\bf 24}$.
\item The two Higgs doublets of the MSSM should mostly be made out of a
pair of Higgs quintet and anti-quintet, which couple to the third
generation.
\end{enumerate}

In the following we discuss two versions of the all-order finite
model.    The model of ref.~\cite{zoup-finite1}, which
will be labeled ${\bf A}$, and a slight variation of this model
(labeled ${\bf B}$), which can also be obtained from the class of the
models suggested in ref.~\cite{zoup-avdeev1} with a modification to
suppress non-diagonal anomalous dimensions.

The  superpotential which describes the two models 
takes the form \cite{zoup-finite1,zoup-kkmz1}
\bea
W &=& \sum_{i=1}^{3}\,[~\frac{1}{2}g_{i}^{u}
\,{\bf 10}_i{\bf 10}_i H_{i}+
g_{i}^{d}\,{\bf 10}_i \overline{\bf 5}_{i}\,
\overline{H}_{i}~] \nn \\\nn
&+&g_{23}^{u}\,{\bf 10}_2{\bf 10}_3 H_{4} 
  +g_{23}^{d}\,{\bf 10}_2 \overline{\bf 5}_{3}\,
\overline{H}_{4}+
g_{32}^{d}\,{\bf 10}_3 \overline{\bf 5}_{2}\,
\overline{H}_{4} \\
&+&\sum_{a=1}^{4}g_{a}^{f}\,H_{a}\, 
{\bf 24}\,\overline{H}_{a}+
\frac{g^{\lambda}}{3}\,({\bf 24})^3~,
\label{zoup-super}
\eea
where 
$H_{a}$ and $\overline{H}_{a}~~(a=1,\dots,4)$
stand for the Higgs quintets and anti-quintets.

The non-degenerate and isolated solutions to $\gamma^{(1)}_{i}=0$ for
the models $\{ {\bf A}~,~{\bf B} \}$ are: 
\bea (g_{1}^{u})^2
&=&\{\frac{8}{5},\frac{8}{5} \}g^2~, ~(g_{1}^{d})^2
=\{\frac{6}{5},\frac{6}{5}\}g^2~,~\nn\\
(g_{2}^{u})^2&=&(g_{3}^{u})^2=\{\frac{8}{5},\frac{4}{5}\}g^2~,\label{zoup-SOL5}\\
(g_{2}^{d})^2 &=&(g_{3}^{d})^2=\{\frac{6}{5},\frac{3}{5}\}g^2~,~\nn\\
(g_{23}^{u})^2 &=&\{0,\frac{4}{5}\}g^2~,~
(g_{23}^{d})^2=(g_{32}^{d})^2=\{0,\frac{3}{5}\}g^2~,
\nonumber\\
(g^{\lambda})^2 &=&\frac{15}{7}g^2~,~ (g_{2}^{f})^2
=(g_{3}^{f})^2=\{0,\frac{1}{2}\}g^2~,~ \nn\\
(g_{1}^{f})^2&=&0~,~
(g_{4}^{f})^2=\{1,0\}g^2~.\nonumber 
\eea 
According to the theorem of
ref.~\cite{zoup-lucchesi1} these models are finite to all orders.  After the
reduction of couplings the symmetry of $W$ is enhanced
\cite{zoup-finite1,zoup-kkmz1}.

The main difference of the models ${\bf A}$ and ${\bf B}$ is that
three pairs of Higgs quintets and anti-quintets couple to the ${\bf
  24}$ for ${\bf B}$ so that it is not necessary to mix them with
$H_{4}$ and $\overline{H}_{4}$ in order to achieve the triplet-doublet
splitting after the symmetry breaking of $SU(5)$.

In the dimensionful sector, the sum rule gives us the following
boundary conditions at the GUT scale \cite{zoup-kkmz1}:
\bea
m^{2}_{H_u}+
2  m^{2}_{{\bf 10}} &=&
m^{2}_{H_d}+ m^{2}_{\overline{{\bf 5}}}+
m^{2}_{{\bf 10}}=M^2~~\mbox{for}~~{\bf A} ~;\\
m^{2}_{H_u}+
2  m^{2}_{{\bf 10}} &=&M^2~,~
m^{2}_{H_d}-2m^{2}_{{\bf 10}}=-\frac{M^2}{3}~,~\nonumber\\
m^{2}_{\overline{{\bf 5}}}+
3m^{2}_{{\bf 10}}&=&\frac{4M^2}{3}~~~\mbox{for}~~{\bf B},
\eea
where we use as  free parameters 
$m_{\overline{{\bf 5}}}\equiv m_{\overline{{\bf 5}}_3}$ and 
$m_{{\bf 10}}\equiv m_{{\bf 10}_3}$
for the model ${\bf A}$, and 
$m_{{\bf 10}}\equiv m_{{\bf 10}_3}$  for ${\bf B}$, in addition to $M$.

\section{PREDICTIONS OF LOW ENERGY PARAMETERS}
 
Since the gauge symmetry is spontaneously broken below $M_{\rm GUT}$,
the finiteness conditions do not restrict the renormalization properties
at low energies, and all it remains are boundary conditions on the
gauge and Yukawa couplings (\ref{zoup-SOL5}), the $h=-MC$ relation,
and the soft scalar-mass sum rule (\ref{zoup-sumr}) at $M_{\rm GUT}$,
as applied in the two models.  Thus we examine the evolution of
these parameters according to their RGEs up
to two-loops for dimensionless parameters and at one-loop for
dimensionful ones with the relevant boundary conditions.  Below
$M_{\rm GUT}$ their evolution is assumed to be governed by the MSSM.
We further assume a unique supersymmetry breaking scale $M_{s}$ (which
we define as the geometric mean of the stop masses) and
therefore below that scale the  effective theory is just the SM.

\begin{figure}
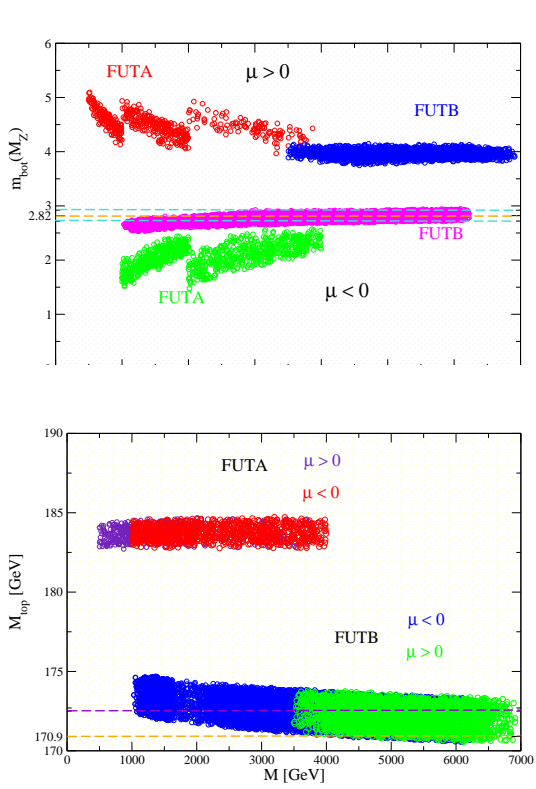

\vspace{0.5cm}
           \centerline{\includegraphics[width=7cm,angle=0]{MvsMBOT.eps}}
\vspace{0.3cm}
           \centerline{\includegraphics[width=7cm,angle=0]{MvsMTOP.eps}}
\vspace{0.5cm}
       \caption{The bottom quark mass at the $Z$~boson scale (upper) 
                and top quark pole mass (lower plot) are shown 
                as function of $M$ for both models.}
\label{fig:MtopbotvsM}
\vspace{-0.5em}
\end{figure}

We now present the comparison of the predictions of the four models with
the experimental data, see ref.~\cite{thefullthing} for more details,
starting with the heavy quark masses. 
In fig.\ref{fig:MtopbotvsM} we show the {\bf FUTA} and {\bf FUTB}
predictions for $M_{\rm top}$ and $m_{\rm bot}(M_Z)$ as a function of the 
unified gaugino mass $M$, for the two cases
$\mu <0$ and $\mu >0$.
In the value of the bottom mass $m_{\rm bot}$,
we have included the corrections coming from bottom
squark-gluino loops and top squark-chargino loops~\cite{deltab}.  We
give the predictions for the running bottom quark mass evaluated at
$M_Z$, $m_{\rm bot}(M_Z) = 2.825 \pm 0.1$~\cite{jens}, to avoid the large
QCD uncertainties inherent for the pole mass.  The value of $m_{\rm bot}$ depends
strongly on the sign of $\mu$ due to the above mentioned 
radiative corrections.  For both models $\bf A$ and $\bf B$ the values
for $\mu >0$ are above the central experimental value, with 
$m_{\rm bot}(M_Z) \sim 4.0 - 5.0$~GeV.  
For $\mu < 0$, on the other hand, model $\bf B$ has
overlap with the experimental allowed values, 
$m_{\rm bot}(M_Z) \sim 2.5-2.8$~GeV,
whereas for model $\bf A$, $m_{\rm bot}(M_Z) \sim 1.5 - 2.6$~GeV, there is only a
small region of allowed parameter space at two sigma level, and only
for large values of $M$. This clearly selects the negative sign of
$\mu$. 

The predictions for the top quark mass $M_{\rm top}$ are $\sim 183$ and
$\sim 172$~GeV in the models ${\bf A}$ and ${\bf B}$
respectively, as shown in the lower plot of fig.~\ref{fig:MtopbotvsM}. 
Comparing these predictions with the most recent experimental value 
$M_{\rm top}^{exp} = (170.9 \pm 1.8)$~GeV~\cite{mt1709}, and recalling
that the theoretical values for $M_{\rm top}$ may suffer from a
correction of $\sim 4 \%$ \cite{zoup-acta}, we see that clearly model
${\bf B}$ is singled out. 
In addition the value of $\tan \beta$ is found to be $\tan \beta \sim 54$ and
$\sim 48$ for models ${\bf A}$ and ${\bf B}$, respectively.
Thus the comparison of the model predictions
with the experimental data is survived only by {\bf FUTB} with $\mu < 0$.

We now analyze the impact of further low-energy observables on the model
{\bf FUTB} with $\mu < 0$. In the case where
all the soft scalar masses are universal at the unfication scale,
there is no region of $M$ below ${\cal O}$(few~TeV) in which $m_{\tilde \tau} >
m_{\chi^0}$ is satisfied (where $m_{\tilde \tau}$ is the lightest $\tilde
\tau$ mass, and $m_{\chi^0}$ the lightest neutralino mass).  But once the
universality condition is relaxed this problem can be solved
naturally, thanks to the sum rule (\ref{zoup-sumr}).  Using this equation
and imposing the conditions of (a) successful radiative
electroweak symmetry breaking, (b) $m_{\tilde\tau}^2>0$ and (c)
$m_{\tilde\tau}> m_{\chi^0}$, a comfortable parameter space is found for 
{\bf FUTB} with $\mu < 0$ (and also for {\bf FUTA} and both signs of $\mu$).

As  additional constraints we consider the following observables: 
the rare $b$~decays $\br(b \to s \gamma)$ and $\br(B_s \to \mu^+ \mu^-)$, 
the lightest Higgs boson mass 
as well as the density of cold dark matter in the Universe, assuming it
consists mainly of neutralinos. More details and a complete set of
references can be found in ref.~\cite{thefullthing}.

For the branching ratio $\br(b \to s \gamma)$, we take
the present 
experimental value estimated by the Heavy Flavour Averaging
Group (HFAG) is~\cite{bsgexp} 
\beq 
\br(b \to s \gamma ) = (3.55 \pm 0.24 {}^{+0.09}_{-0.10} \pm 0.03) 
                       \times 10^{-4},
\label{bsgaexp}
\eeq
where the first error is the combined statistical and uncorrelated systematic 
uncertainty, the latter two errors are correlated systematic theoretical
uncertainties and corrections respectively. 

For the branching ratio $\br(B_s \to \mu^+ \mu^-)$, the SM prediction is
at the level of $10^{-9}$, while the present
experimental upper limit from the Tevatron is 
$5.8 \times 10^{-8}$ at the $95\%$ C.L.~\cite{bsmmexp}, providing the
possibility for the MSSM to dominate the SM contribution.

Concerning the lightest Higgs boson mass, $M_h$, the SM bound of
$114.4$~GeV~\cite{LEPHiggsSM_MSSM} can be used. For the
prediction we use the code 
{\tt FeynHiggs}~\cite{feynhiggs,mhiggslong,mhiggsAEC,mhcMSSMlong}. 

The lightest supersymmetric particle (LSP) is an
excellent candidate for cold dark matter (CDM)~\cite{EHNOS}, with a density 
that falls naturally within the range 
\beq
0.094 < \Omega_{\rm CDM} h^2 < 0.129
\label{cdmexp}
\eeq
favoured by a joint analysis of WMAP and other astrophysical and
cosmological data~\cite{WMAP}.

\begin{figure}
           \centerline{\includegraphics[width=7cm,angle=0]{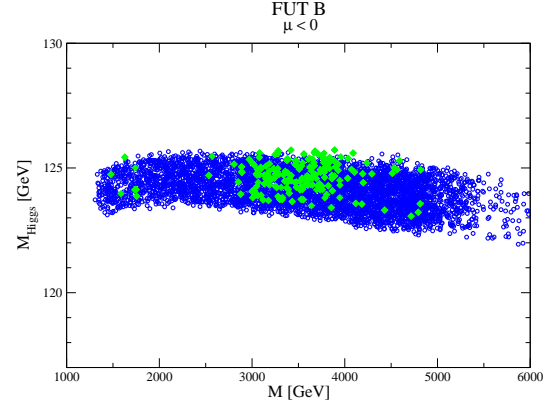}}
        \caption{The lightest Higgs mass, $M_h$,  as function of $M$ for
          the model {\bf FUTB} with $\mu < 0$, see text.}
\label{fig:Higgs}
\vspace{-0.5em}
\end{figure}

The prediction for $M_h$ of {\bf FUTB} with $\mu < 0$ is shown in
fig.~\ref{fig:Higgs}. 
The constraints from the two $B$~physics observables are taken into
account. In addition the CDM constraint (evaluated with 
{\tt Micromegas}~\cite{micromegas}) is fulfilled for the lighter
(green) points in the plot, see ref.~\cite{thefullthing} for details. 
The lightest Higgs mass ranges in 
\beq
M_h \sim 121-126~{\rm GeV} , 
\label{eq:Mhpred}
\eeq 
where the uncertainty comes from
variations of the soft scalar masses, and
from finite (i.e.~not logarithmically divergent) corrections in
changing renormalization scheme.  To this value one has to add $\pm 3$
GeV coming from unkonwn higher order corrections~\cite{mhiggsAEC}. 
We have also included a small variation,
due to threshold corrections at the GUT scale, of up to $5 \%$ of the
FUT boundary conditions.  
Thus, taking into account the $B$~physics constraints (and possibly the
CDM constraints) results naturally in a light Higgs boson that fulfills
the LEP bounds~\cite{LEPHiggsSM_MSSM}. 

In the same way the whole SUSY particle spectrum can be derived. 
The resulting SUSY masses for {\bf FUTB} with $\mu < 0$ are rather large.
The lightest SUSY particle starts around 500 GeV, with the rest
of the spectrum being very heavy. The observation of SUSY particles at
the LHC or the ILC will only be possible in very favorable parts of the
parameter space. For most parameter combination only a SM-like light
Higgs boson in the range of eq.~(\ref{eq:Mhpred}) can be observed.

We note that with such a heavy SUSY spectrum 
the anomalous magnetic moment of the muon, \mbox{$(g-2)_\mu$}
(with $\amu \equiv (g-2)_\mu/2$), gives only a negligible correction to
the SM prediction.  
The comparison of the experimental result and the SM value~\cite{DDDD}
\beq
\amuexp-\amutheo = (27.5 \pm 8.4) \times 10^{-10}.
\label{delamu}
\eeq
would disfavor {\bf FUTB} with $\mu < 0$ by about $3\,\sigma$. However, 
since the SM is not regarded as excluded by $(g-2)_\mu$, 
we still see {\bf FUTB} with $\mu < 0$ as the only surviving model.
A more detailed numerical analysis, also using
{\tt Suspect}~\cite{suspect} for the RGE
running, and including all theory uncertainties for the models will be
presented in ref.~\cite{thefullthing}.

\bigskip
We acknowledge very useful conversations with\\ G.~Belanger, F.~Boudjema and
A.~Djouadi. 
This work is supported by the EPEAEK programme ``Pythagoras'' of the
Greek Ministry of Education and co-founded by the European Union
(75\%) and the Hellenic State (25\%), and  by a
PAPIIT-UNAM IN115207 grant.
Work supported in part by the European
Commission under the Research and Training
Network contract MRTN-CT-2006-035505.
\\[-2.5em]

\end{document}